\documentclass[pra,preprint,superscriptaddress,amsmath,amssymb,floatfix]{revtex4-1}

\usepackage{graphicx}
\usepackage{wrapfig}

\usepackage[tableposition=top]{caption}

\begin{document} 
\title{Density functional theory study of vacancy induced magnetism in Li$_{3}$N}
\author{A. \"Ostlin}
\affiliation{Theoretical Physics III, Center for Electronic
Correlations and Magnetism, Institute of Physics, University of
Augsburg, D-86135 Augsburg, Germany}
\author{L. Chioncel}
\affiliation{Augsburg Center for Innovative Technologies, University of Augsburg,
D-86135 Augsburg, Germany}
\affiliation{Theoretical Physics III, Center for Electronic
Correlations and Magnetism, Institute of Physics, 
University of
Augsburg, D-86135 Augsburg, Germany} 
\author{E. Burzo}
\affiliation{Faculty of Physics, Babes-Bolyai University 40084 Cluj-Napoca, Romania}
 
\begin{abstract} 
The effect of lithium vacancies in the hexagonal structure of 
$\alpha-$Li$_3$N, is studied within the framework of density 
functional theory. Vacancies ($\square$) substituting for 
lithium in $\alpha$-Li$_3$N
are treated within the coherent 
potential approximation as alloy components. According to our results
long range N($p$)-ferromagnetism ($\sim 1$ $\mu_B$) sets in for vacancy substitution 
within the [Li$_2$N] layers ($x \ge 0.7$)  with no significant change in
unit cell dimensions. By total energies differences we established that
in-plane exchange couplings are dominant. Vacancies substituting 
inter-plane Li, leads to a considerable structural
collapse ($c/a \approx 0.7$) and no magnetic moment formation.
\end{abstract} 

\maketitle 

\section{Introduction}\label{sec_intro}
\begin{figure}[h!]
\centering
\includegraphics[scale=0.7]{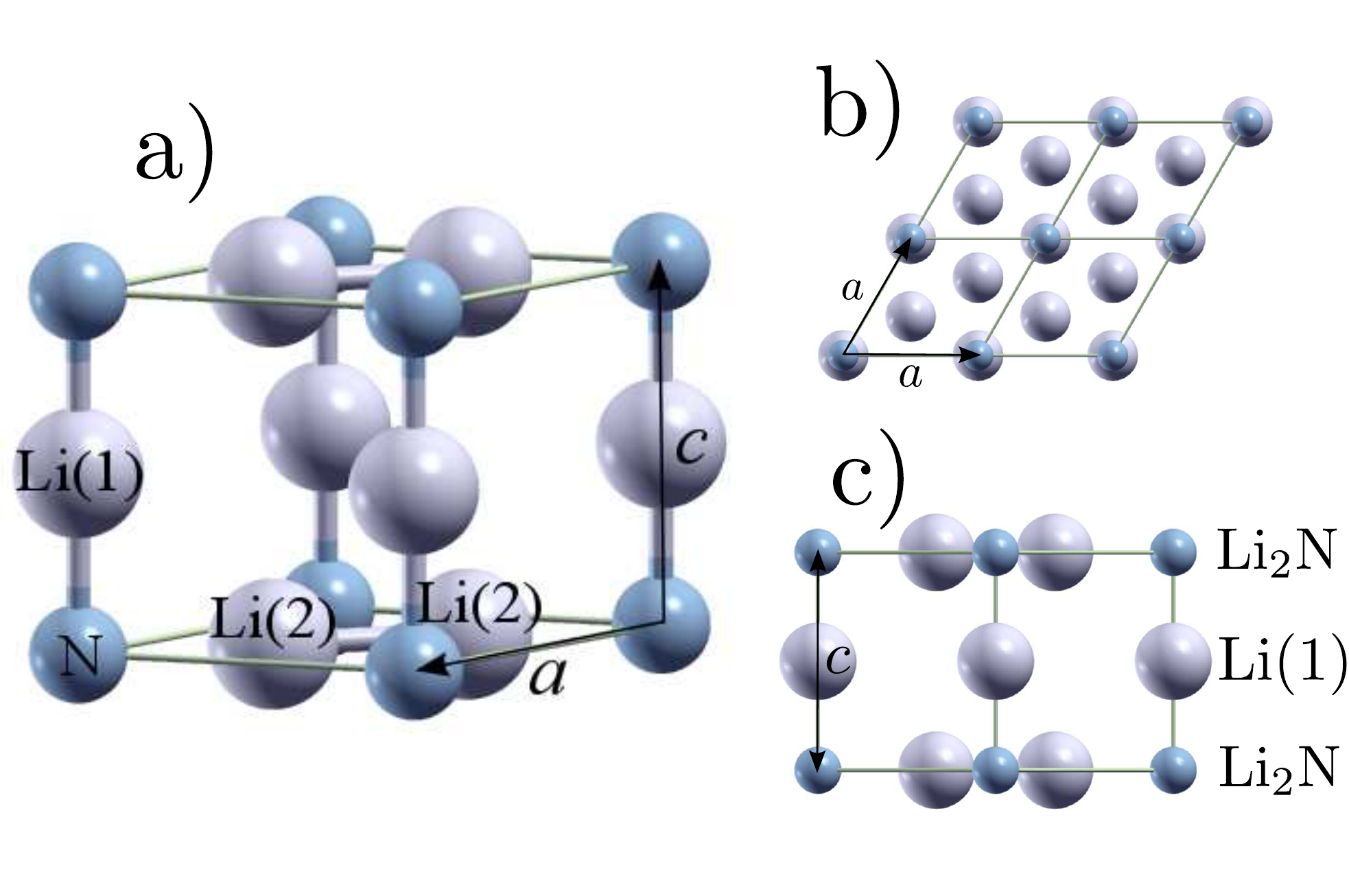}
\caption{Crystal structure of $\alpha-$Li$_3$N. (a) The crystal is made of hexagonally close-packed layers of Li$_2$N, where the Li atoms are denoted by Li(2). These layers are alternated by Li layers, where the Li is denoted by Li(1). The Li(1) atom is placed in between the N in the Li$_2$N layers. (b) Top view of the Li$_2$N layer. (c) Side view showing the alternating stacking of the Li$_2$N and Li(1) layers.}\label{crys}
\end{figure}

Lithium nitride (Li$_3$N) is the only known stable alkali metal nitride. 
At ambient conditions it crystallizes in the hexagonal $P6/mmm$ structure ($\alpha-$Li$_3$N)
seen in Fig.~\ref{crys}. The structure consists of  
alternating Li and Li$_2$N layers~\cite{ra.sc.76,rabe.82}. 
The basal $ab$-plane contains edge sharing planar Li hexagons 
centred by nitrogen. These planes are connected through a further lithium 
ion along the $c$-axis located directly on top of the N. 
Each nitrogen is coordinated by eight lithium atoms in a hexagonal bipyramid geometry. 
Within the Li$_2$N layers ($ab$-planes) the atoms are close-packed, while the Li layers 
form an open structure. 
In the following the Li above the N atom will be denoted as Li(1) and the Li atoms within 
the Li$_2$N layers will be denoted as Li(2), see Fig.~\ref{crys}. It is known that the bonding in 
this material is mainly ionic, with the three lithium atoms donating their 
$2s$ electrons to the nitrogen, resulting in Li$^+$ ions and a N$^{3-}$ 
ion~\cite{sc.sc.78}. Lithium nitride is a common material used in the solid-state
battery technology~\cite{rabe.82,ta.ar.01}: the indirect band gap is about $2.1$ eV~\cite{br.bl.79} 
and the material has exceptionally high Li$^{+}$ ion mobility~\cite{rabe.82}, which arises from 
the cationic vacancies within the [Li$_2$N] planes.

Sachsze and Juza~\cite{sa.ju.49} found that the Li(1) atom could easily be 
substituted by a transition metal atom. This possibility has recently generated 
a substantial amount of interest due to the large magnetic moment observed at 
the substituted site~\cite{kl.sc.02,no.wa.02,an.an.14,je.mc.14,je.ke.15}. This 
magnetic moment supersedes that of the donor transition metal element and is 
attributed to the absence of orbital moment quenching and to on-site Coulomb interactions~\cite{an.an.14}. 
These findings have led to predictions that transition 
metal doped Li$_3$N might be considered for replacing rare-earth permanent 
magnets~\cite{je.mc.14}. The magnetic behaviour associated with 
the presence of transition metal atoms lead us to raise the question whether magnetism 
is still triggered by non-magnetic atoms, or in the extreme case by vacancies, that are used as
dopants in the structure of lithium nitride. In the present paper the effect 
of vacancy induced magnetism is studied theoretically using density functional 
theory (DFT)~\cite{ho.ko.64,ko.sh.65,kohn.99} methods.

The electronic structure of the ordered $\alpha-$Li$_3$N crystal has been studied using 
density functional~\cite{kerk.81,bl.re.84,bl.sc.85,sa.sc.96} and Hartree-Fock~\cite{do.pi.84,se.pa.91} theories. 
Similar studies have been performed on the Li ion vacated structure~\cite{wu.do.09}, 
also using molecular dynamics~\cite{sa.sc.96}. 
None of these calculations took the spin-polarization of the charge density into account, 
and therefore magnetic effects were not reported previously.

Besides $\alpha-$Li$_3$N, two other polymorphs exist: the hexagonal $\beta-$Li$_3$N 
stable at moderate pressure (4.2 kbar and 
300 K)~\cite{be.ju.88,la.ma.05} and the cubic $\gamma-$Li$_3$N stable up to 
$200$ GPa~\cite{la.ma.05,la.yo.08}. It has been demonstrated that Li$_3$N 
retains its ionic character through the significant structural transformation~\cite{la.ma.05}. 
Upon pressure, the band gap is increased across the hexagonal ($\beta$) to cubic 
($\gamma$) structural transition. The increase in the magnitude of the band gap 
is a consequence of the loss of the lowest laying conduction 
band as one goes from $\beta\rightarrow\gamma$~\cite{la.yo.08}.
The large ($5.5$ eV) gap $\gamma-$Li$_3$N material shows an anomalous color,
which could be due to the presence of vacancies (color centres)~\cite{la.yo.08}.
It is commonly found that the lithium vacancy concentration increases 
with doping, especially at the Li(2) sites~\cite{wu.do.11}. 
In the extreme case of LiNiN, where full substitution of the Li(1) 
atom with nickel has been achieved, one of the Li(2) sites is completely vacated, 
creating a crystal structure with the $P\bar{6}m2$ space group~\cite{ba.bl.99,st.go.04}.
As expected, the presence of Ni determines the magnetic properties of LiNiN. 
On the other hand, in the absence of a transition metal, it is obvious to ask  
if $p$-orbital magnetism can be induced due to vacancies in this system. 

Magnetism is most often associated with systems containing partially occupied $d-$ or $f-$shells.
In transition metals or rare earth elements, Hund's exchange energy and the kinetic energy
are of similar magnitude, and magnetism may appear if the Hund's energy is large enough~\cite{mohn_book}.
A similar discussion has been extended to $p-$electronic systems, containing C, N or O~\cite{vo.bo.10}.
In order to elucidate on the effect of vacancies on total energy and magnetism, we have in this 
paper performed a DFT study on the electronic structure of Li$_3$N with Li atom vacancies.
In our work we show that the creation of Li atom vacancies within the perfect $\alpha-$Li$_3$N
structure induces: (i) a volume collapse of a magnitude which depends on the symmetry position of 
the removed Li atom, and also (ii) magnetism, with moments centred on the N ions.
In the case of a removal of the Li(1) atom, a large structural collapse occurs which leads to
a vanishing magnetic moment. For a removal of Li(2) atoms the volume reduction is not 
drastic, and a magnetic moment is still formed. We also compute the vacancy concentration needed
to form long-range magnetic ordering between the vacancy-induced N-moments in this system.

The paper is organized as follows: In Section \ref{sec_compdet} we introduce
the theoretical tools, and present the computational details. 
Section~\ref{sec_resdis} presents the calculated ground-state lattice 
parameters of Li$_3$N, and the effect of a complete removal of inter- and 
in-plane Li atoms. We also investigate the partial, non-stoichiometric, replacement 
of one of the in-plane Li atoms by a vacancy, employing the coherent potential 
approximation. In addition the band 
structure of Li$_3$N, with and without vacancies, and the
 ordering of the magnetic moments is discussed. Section~\ref{sec_discus} provides 
a discussion, and Section~\ref{sec_conc} gives a conclusion.

\section{Methodology and Computational details}\label{sec_compdet}
The electronic structures for all systems were obtained by Kohn-Sham density functional theory~\cite{ho.ko.64,ko.sh.65}. Three different basis sets to represent the Kohn-Sham orbitals were used, namely the full-potential linearised augmented plane wave (FPLAPW) method as implemented in \emph{elk}~\cite{elk}, the exact muffin-tin orbitals (EMTO) method~\cite{an.je.94.2,vi.sk.00,vito.01} and the linearised muffin-tin (LMTO) method~\cite{ande.75}. 
As seen from Fig.~\ref{crys}, Li$_3$N has an open structure, with large interstitial areas in the Li(1) layer planes. To model this large interstitial region, the LMTO and EMTO methods introduce so called `empty spheres' into
the structure in order to improve upon the spherical potential approximation. The full potential is taken
into account in the FPLAPW method.

To model the non-stoichiometric substitution of Li atoms with vacancies, we employed the coherent potential approximation (CPA) within the EMTO method. 
The EMTO-CPA method together with the full-charge density (FCD)~\cite{ko.vi.00} scheme has proven to be able to accurately determine the total energy due to anisotropic lattice distortions (e.g hexagonal $c/a$-ratios) for real systems with substitutional disorder \cite{vi.ab.01,ti.ye.17}.

For the FPLAPW calculations the plane-wave cutoff $K_{max}$ was set so that $R_{mt}K_{max}=8.0$, where
$R_{mt}$ is the smallest muffin-tin radius. The radii of the muffin tins were set to $R=1.7$ a.u. for N, $R=1.45$ a.u. for Li(1) and $R=1.8$ a.u. for Li(2), and were kept fixed for all volumes and $c/a$-ratios considered. The irreducible Brillouin zone was sampled with 333 $\mathbf{k}$-points, and a Fermi smearing function was employed at room temperature. The N $1s$-state was treated as core, while the remaining states were treated as valence. The exchange correlation potential was treated within the local density approximation (LDA)~\cite{pe.wa.92}.

The EMTO calculations were done with a angular momentum cutoff keeping the $spd$-states. The kink-cancellation equation
was solved along a semicircular contour in the complex energy plane, with the bottom at $-1.5$ Ry below the Fermi level. 
In this way the contour enclosed the N $2s$ semicore states. The irreducible Brillouin zone was sampled 
with 324 $\mathbf{k}$-points. Empty spheres where included at the positions ($\frac{2}{3}$,$\frac{1}{3}$,$\frac{1}{2}$) and
($\frac{1}{3}$,$\frac{2}{3}$,$\frac{1}{2}$), i.e. in the same plane as Li(1) and right above the Li(2), in order to cover
the interstitial volume. The energy dependence of the slope matrix was parametrized using the two-center Taylor expansion
technique~\cite{ki.vi.07}.
The volume vs. $c/a$-ratio total energy contours were calculated for 5 volumes and 5 $c/a$-ratios, and interpolation
and energy minimization were performed to find the minima. 

\section{Results and discussion}
\label{sec_resdis}

\subsection{Perfect Li$_3$N}
\begin{figure}[h!]
\centering
\resizebox*{8cm}{!}{\includegraphics{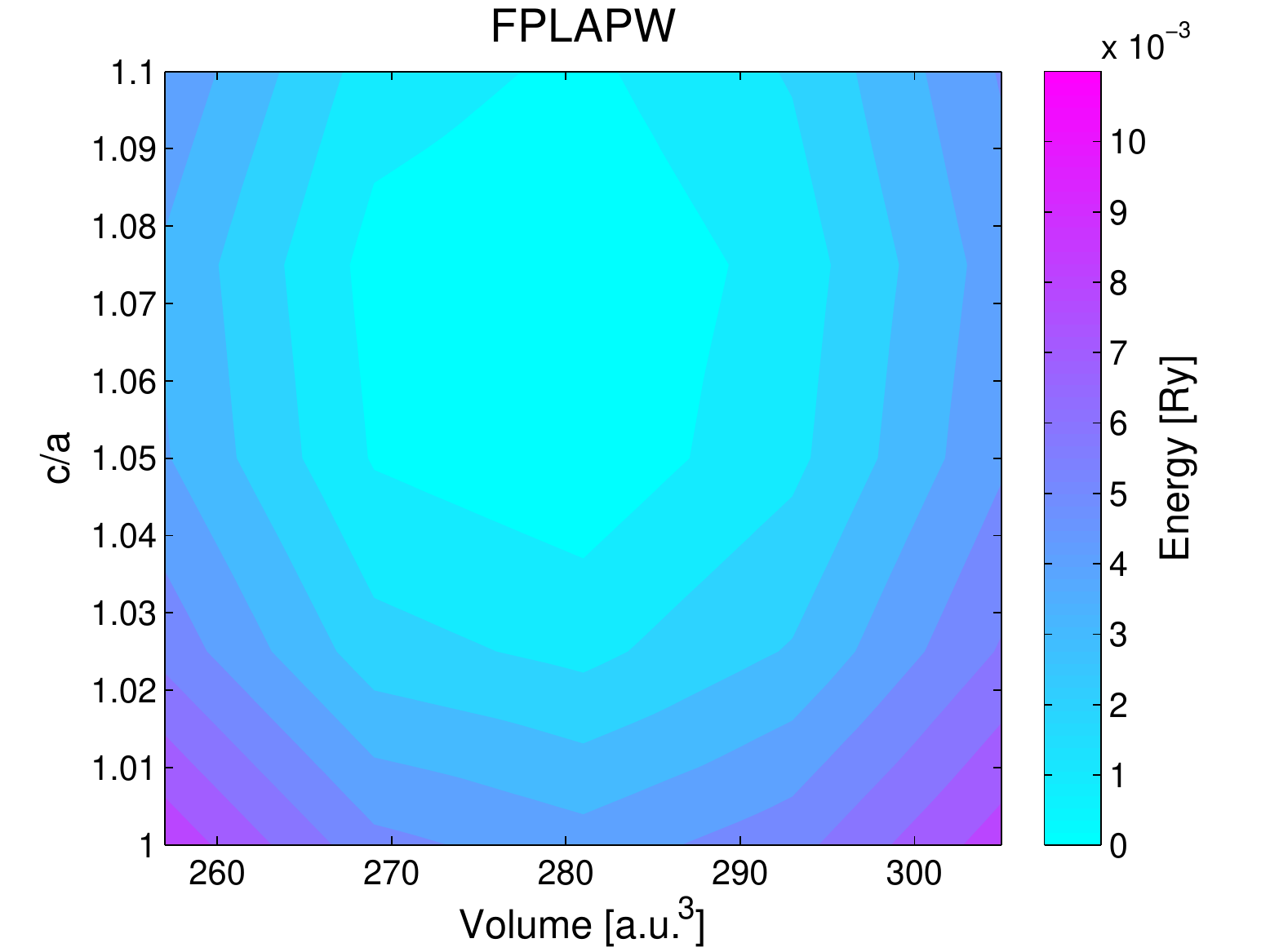}}\resizebox*{8cm}{!}{\includegraphics{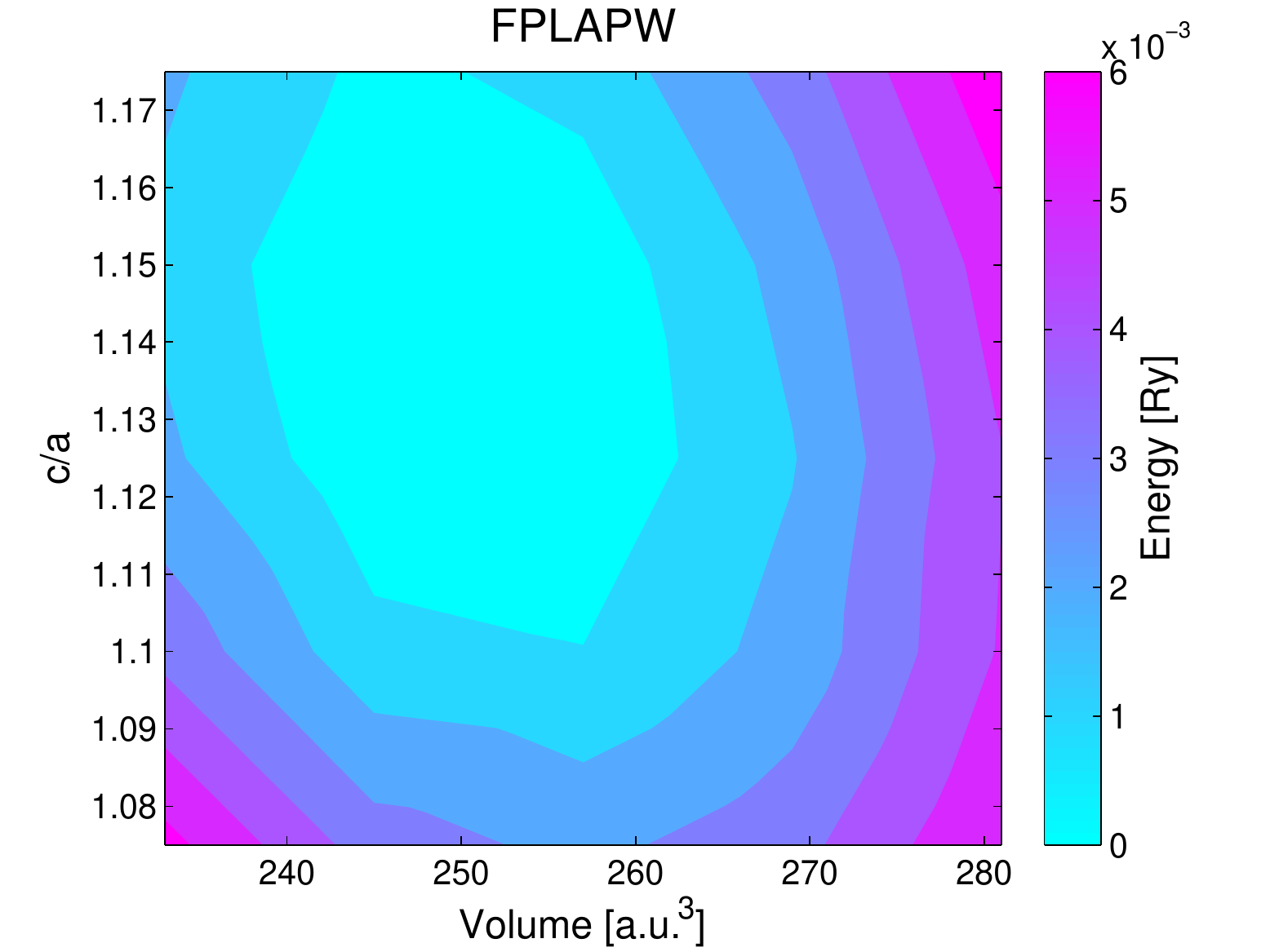}}
\caption{Left: Energy map as a function of volume and $c/a$-ratio for perfect Li$_3$N, calculated with the FPLAPW method. Right: Energy map as a function of volume and $c/a$-ratio for Li(2)$_1\square_1$Li(1)N, where the vacancy $\square$ is of type Li(2)}\label{fig_li3n}
\end{figure}

\begin{figure}[h!]
\centering
\resizebox*{8cm}{!}{\includegraphics{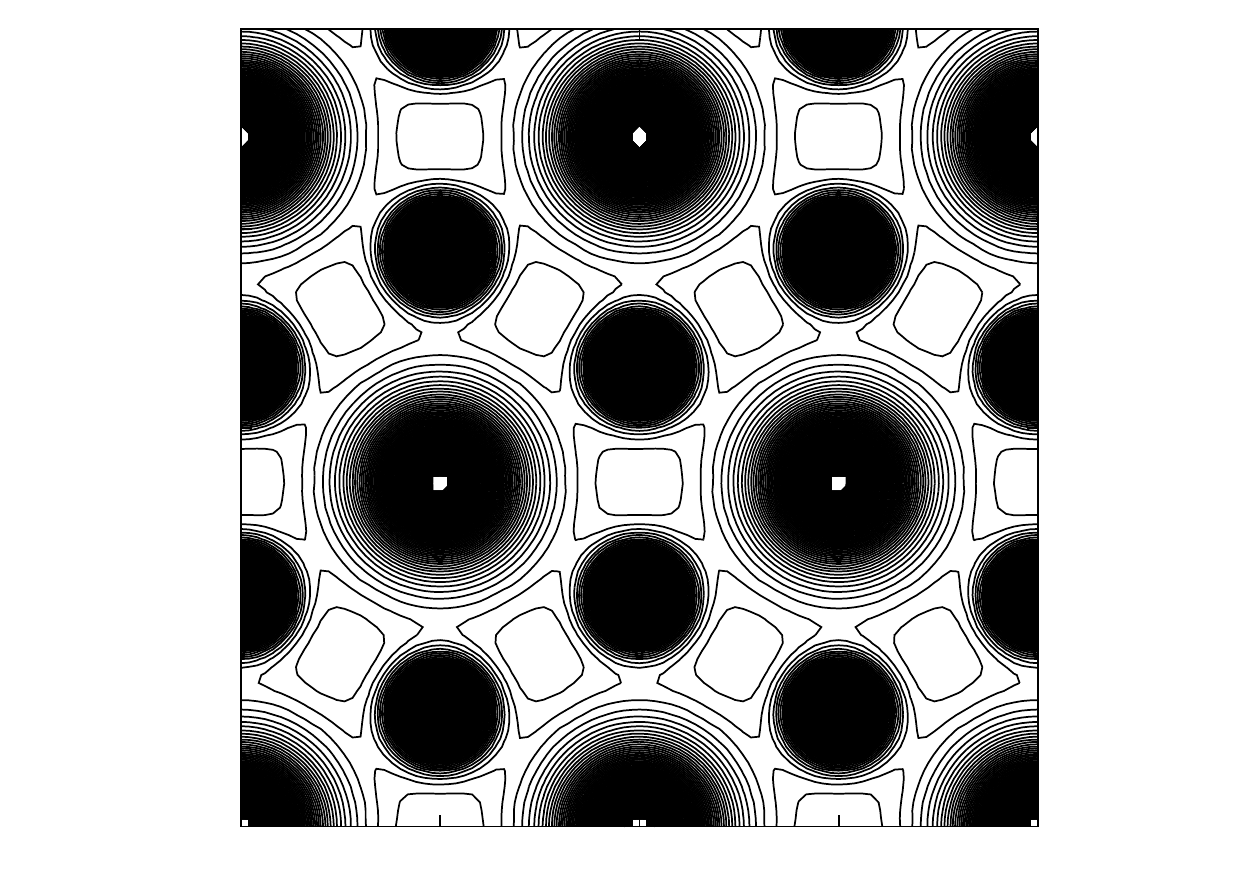}}\resizebox*{8cm}{!}{\includegraphics{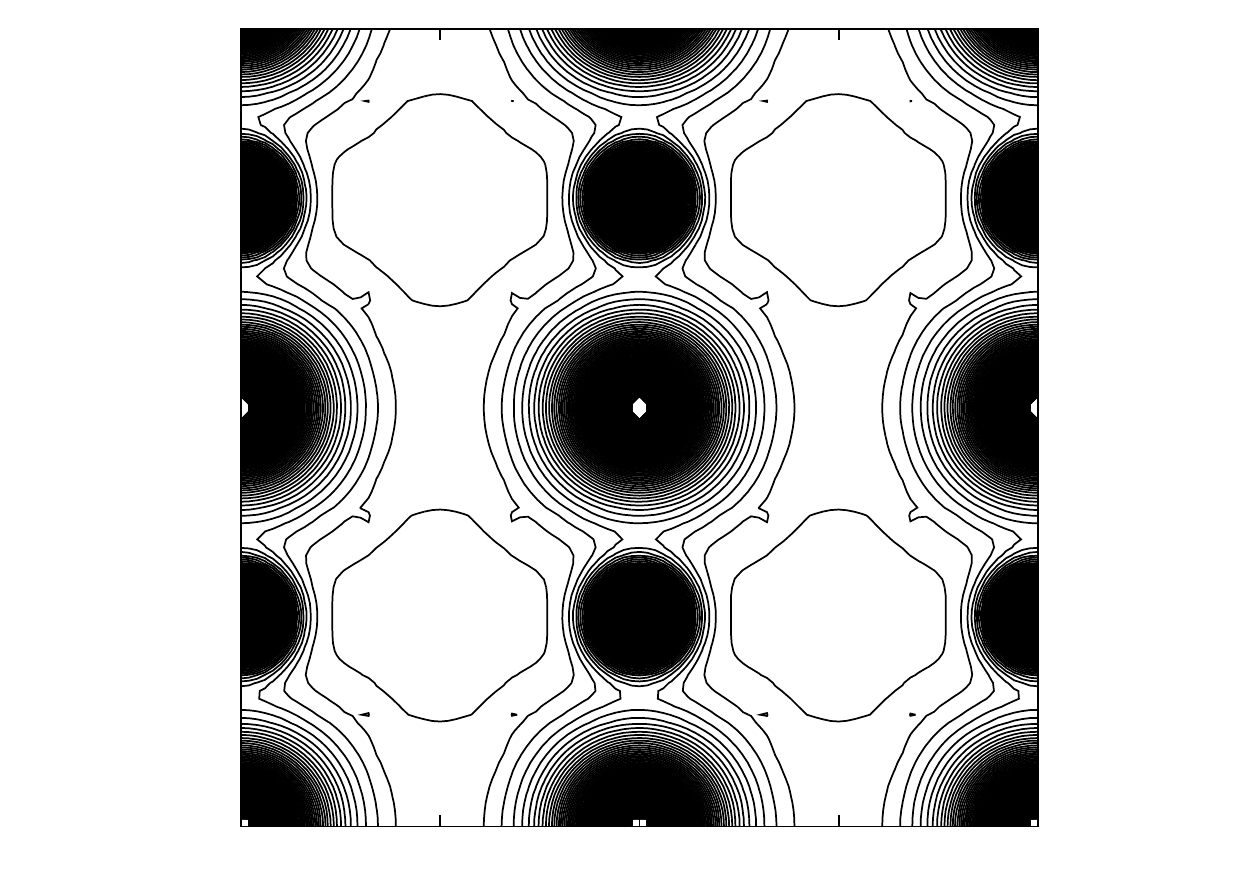}}
\caption{Left: Total charge density for Li$_3$N within the $ab$-plane. Right: Total charge density for Li$_3$N in the (100)-plane. The contour line spacing is 0.01 in atomic units. Calculations performed at volume $V=281$ a.u.$^3$ ($c/a=1.05$).}\label{cont1}
\end{figure}

We first optimized the volume and $c/a$-ratio for the perfect Li$_3$N. The energy landscape as given by the FPLAPW method as a function of volume and $c/a$-ratio can be seen in Fig. \ref{fig_li3n} (left). The FPLAPW method leads to a ground-state volume of $V_0 = 278.12$ a.u.$^3$ and $c/a = 1.070$, while the EMTO method gives a volume of $V_0 = 276.62$ a.u.$^3$  and $c/a = 1.069$. The results are in good agreement with each other, indicating that the spherical approximation is valid.
Lattice optimization was also done within the FPLAPW method employing the PBE-GGA~\cite{pe.bu.96} for the exchange-correlation potential, which results in a slight increase in volume ($6\%$) compared to the LDA result, and $c/a=1.062$.
Previous Hartree-Fock calculations have given values of $V_0 = 292.46$ a.u.$^3$ and $c/a=1.0637$ (Ref.~\cite{do.pi.84}). 
Our results can be seen to agree well with experimental (see Table \ref{tab1}) and previous theoretical studies.
Also note that the calculations predict an insulating state for Li$_3$N, in agreement with experiment. Spin-polarized calculations were also performed, and no magnetic moment was found. Fig. \ref{cont1} shows contour plots of the total charge density of Li$_3$N as calculated within the EMTO method. The charge densities show good agreement compared with previous studies~\cite{kerk.81,la.yo.08}. Note the close packing within the Li$_2$N layer (left) and the large interstitial region between the layers (right).

\begin{table}[h!]
\caption{Volumes in atomic units (first rows) and $c/a$-ratios (second rows) for perfect Li$_3$N, Li(2)$_1\square_1$Li(1)N (in-plane Li(2) vacancy) and for Li(2)$_{1.5}\square_{0.5}$Li(1)N calculated within our methods. Comparison with experimental data is also presented.}
\label{tab1}
\begin{ruledtabular}
\begin{tabular}{cccccc} 
 & & FPLAPW & EMTO & Ref.~\cite{sc.sc.78} & Ref.~\cite{je.mc.14} \\
 \hline
Li$_3$N & $(V_0)$ & 278.12 & 276.62 & 299.98 & 301.65 \\
        & $(c/a)$ &  1.070  & 1.069   & 1.0634 & 1.0597 \\
 \hline
Li$_2$N & $(V_0)$ & 248.50 & 252.31 & & \\
 (inplane) & $(c/a)$ & 1.141   & 1.071   & & \\
 \hline
Li(2)$_{1.5}\square_{0.5}$Li(1)N & $(V_0)$ & & 269.04 & & \\
  & $(c/a)$ & & 1.065 & & \\ 
\end{tabular}
\end{ruledtabular}
\end{table}

\subsection{Vacancies within the [Li$_2$N]-plane}
A vacancy (denoted by $\square$) created by removing one of the (geometrically equivalent) 
Li(2) atoms changes the space group to $P\bar{6}m2$. If the lattice 
parameters from the perfect Li$_3$N are kept, the calculations for 
the Li(2) vacated structure predicts a non-zero density of states 
at the Fermi level, indicating that the system has turned metallic. 
If the spin-symmetry is broken, the system turns ferromagnetic (FM) with an integer 
moment of 1 $\mu_B$.
The underlying reasons for the onset of magnetization will be discussed in more
detail in Sections~\ref{res_bands} and~\ref{res_mag}.
In order to see whether the removal of the Li(2) atom 
has any effect on the geometry of the structure we optimized the lattice 
parameters, see Fig. \ref{fig_li3n} (right). We found that the ground-state volume 
within EMTO reduces slightly to $V_0 = 252.31$ a.u.$^3$, and the $c/a$-ratio 
becomes 1.071. For the FPLAPW, we got $V_0 = 248.50$ a.u.$^3$, and the 
$c/a$-ratio becomes 1.141. The slight discrepancy in the $c/a-$ratio between the FPLAPW
and EMTO methods could possibly be due to the spherical approximation used in the latter
as also reported previously~\cite{oe.vi.11}.

\begin{figure}[h!]
\centering
\resizebox*{8cm}{!}{\includegraphics{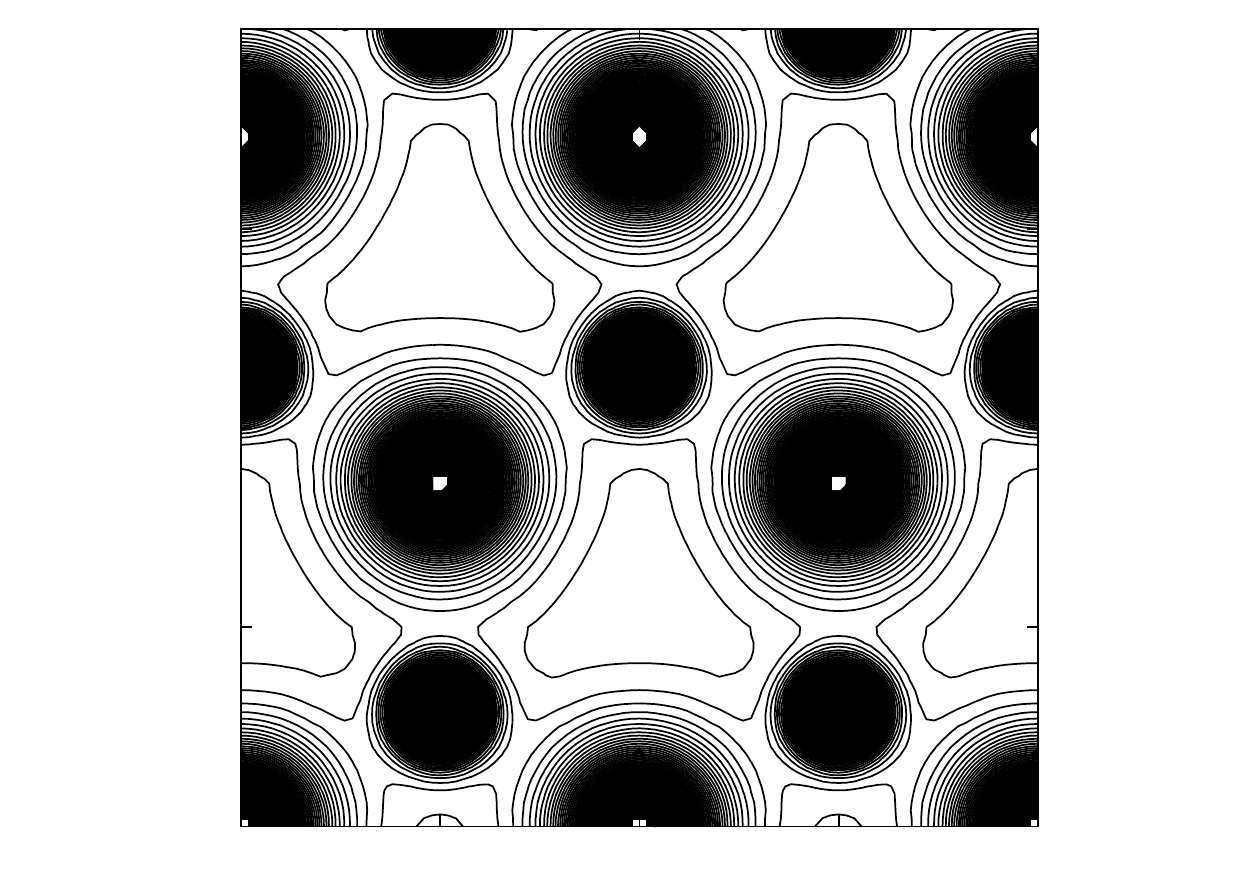}}\resizebox*{6.2cm}{!}{\includegraphics{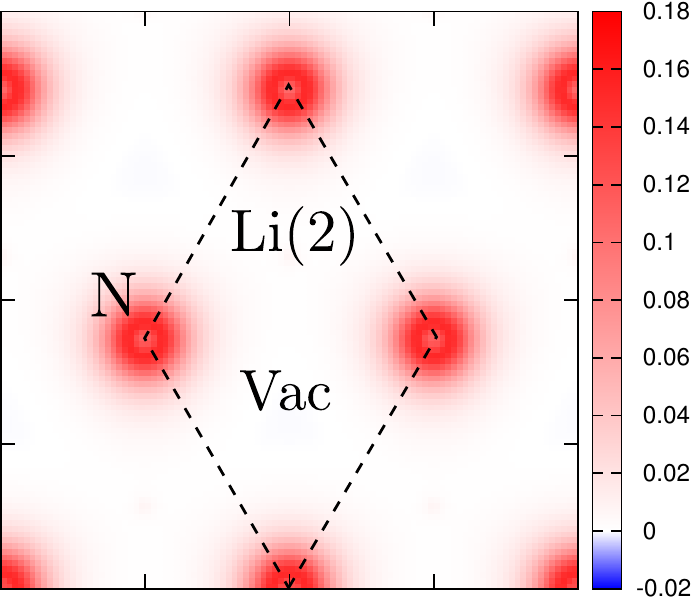}}
\caption{Left: Total charge density $n_{\uparrow}(\mathbf{r})+n_{\downarrow}(\mathbf{r})$ for Li$_2\square$N in the $ab$-plane. Spacing between contour lines is 0.01 in atomic units. Right: Magnetization density $n_{\uparrow}(\mathbf{r})-n_{\downarrow}(\mathbf{r})$ for Li$_2\square$N in the $ab$-plane. Calculations performed at volume $V=281$ a.u.$^3$ ($c/a=1.05$).}\label{cont2}
\end{figure}

Fig. \ref{cont2} shows the total charge density in real space plotted within the Li$_2$N $ab$-plane, where a comparison
with the case of the perfect Li$_3$N structure (Fig. \ref{cont1} (left)) shows the vacant Li(2) positions in 
the lattice. The magnetization density, defined as the difference between the real space majority and minority
spin density, $n_{\uparrow}(\mathbf{r})-n_{\downarrow}(\mathbf{r})$, is shown on the right hand-side of Fig. \ref{cont2}. 
Note that the magnetization density is sizeable mainly around the nitrogen 
ions, indicating that the spin density is highly localized around the nitrogen 
positions in real space.

\begin{figure}[h!]
\includegraphics[scale=1.0]{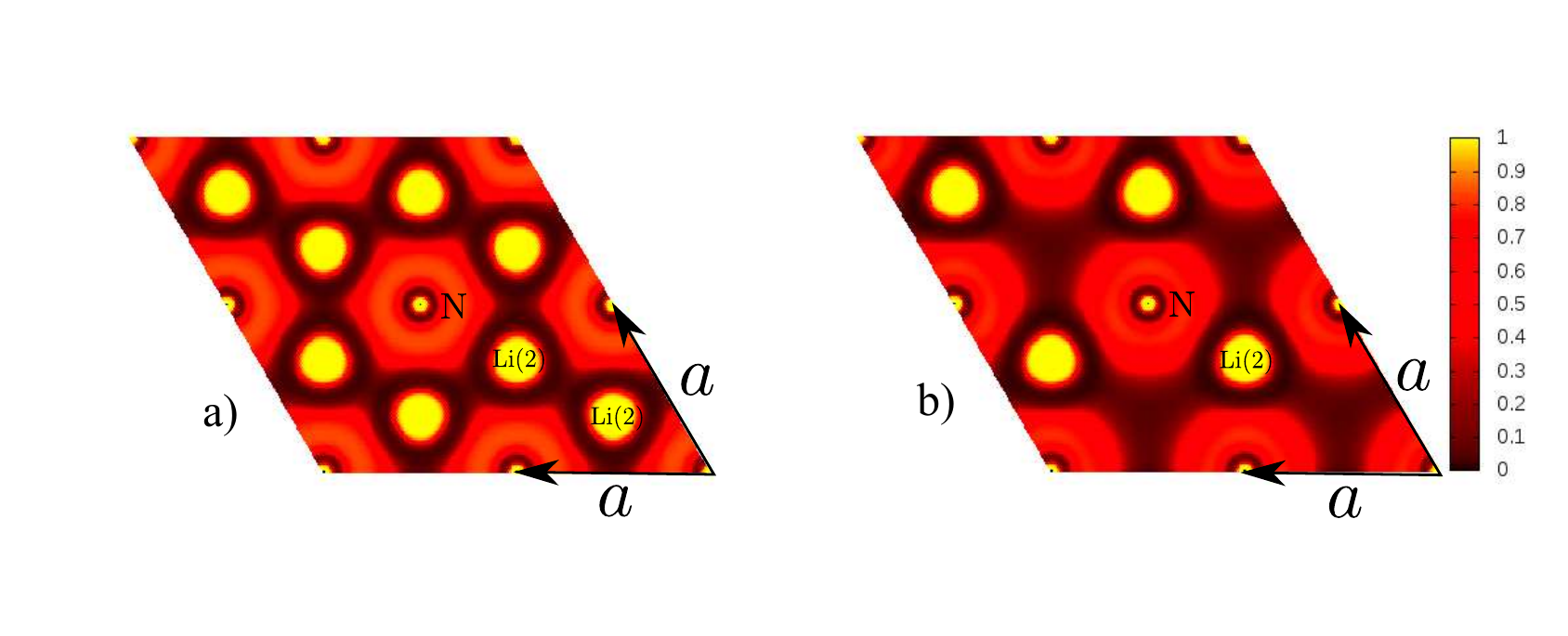}
\caption{Electronic localization function (ELF) for (a) Li$_3$N and (b) Li$_2\square$N in the $ab$-plane. Calculations performed at volume $V=281$ a.u.$^3$ ($c/a=1.05$).}\label{elf}
\end{figure}

The chemical bonding of the pure and the vacated Li$_3$N structure can be visualized by
the electronic localization function (ELF)~\cite{be.ed.90}, presented in Fig.~\ref{elf}. In the case of pure
Li$_3$N (left part of Fig.~\ref{elf}) the bonding is ionic, as found previously~\cite{sc.sc.78,bl.sc.85}. 
Ionic bonding is also seen in the vacated structure (right part of Fig.~\ref{elf}).
In both cases, the ELF shows two maxima as one goes radially outwards from the nitrogen ions, indicating
two shells. The lithium ions only show one maximum, suggesting that the outer $2s$-shell electron has
been mainly transferred to the nitrogen ion. 

\begin{figure}[h!]
\centering
\resizebox*{8cm}{!}{\includegraphics{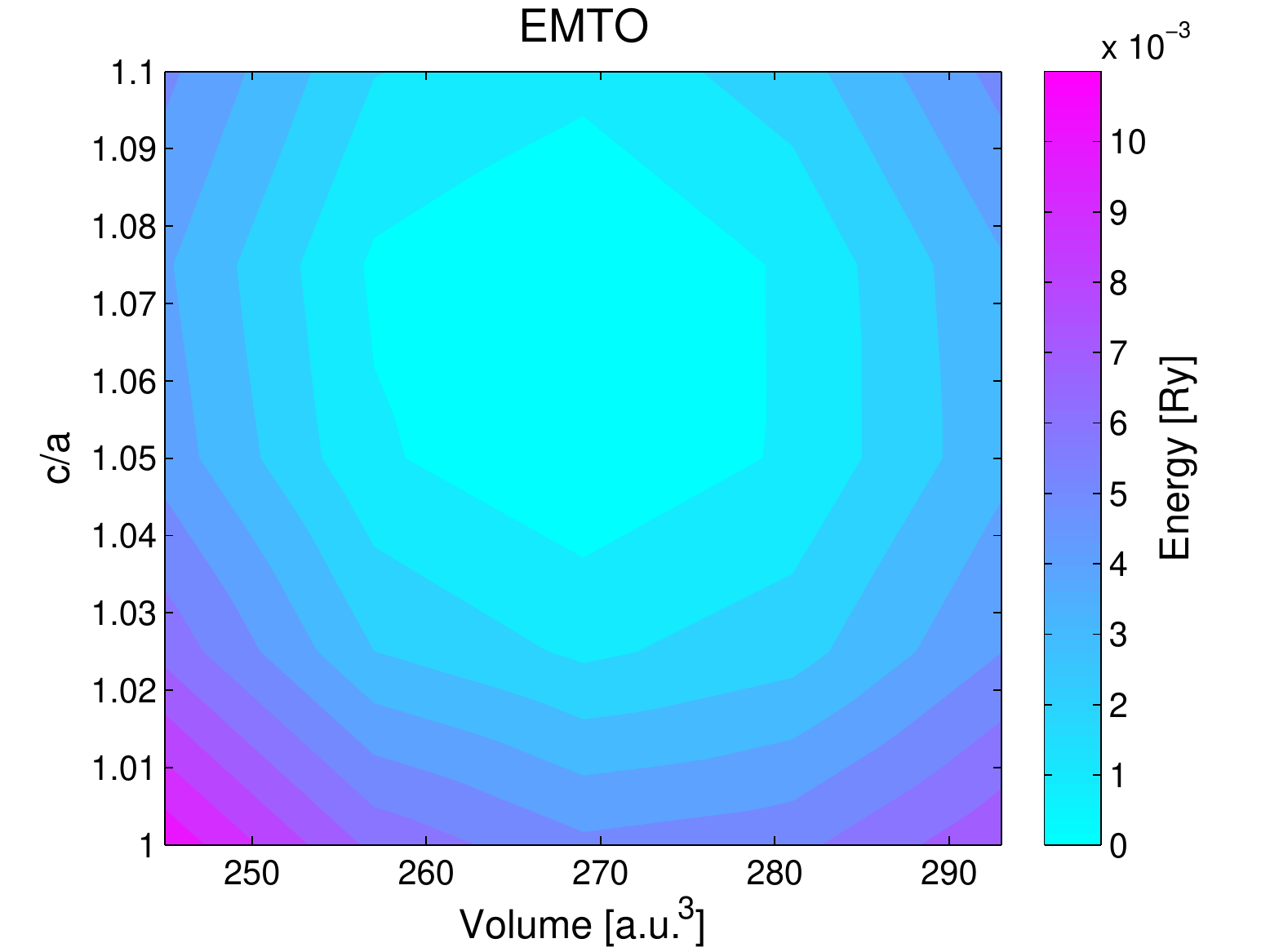}}\resizebox*{8cm}{!}{\includegraphics{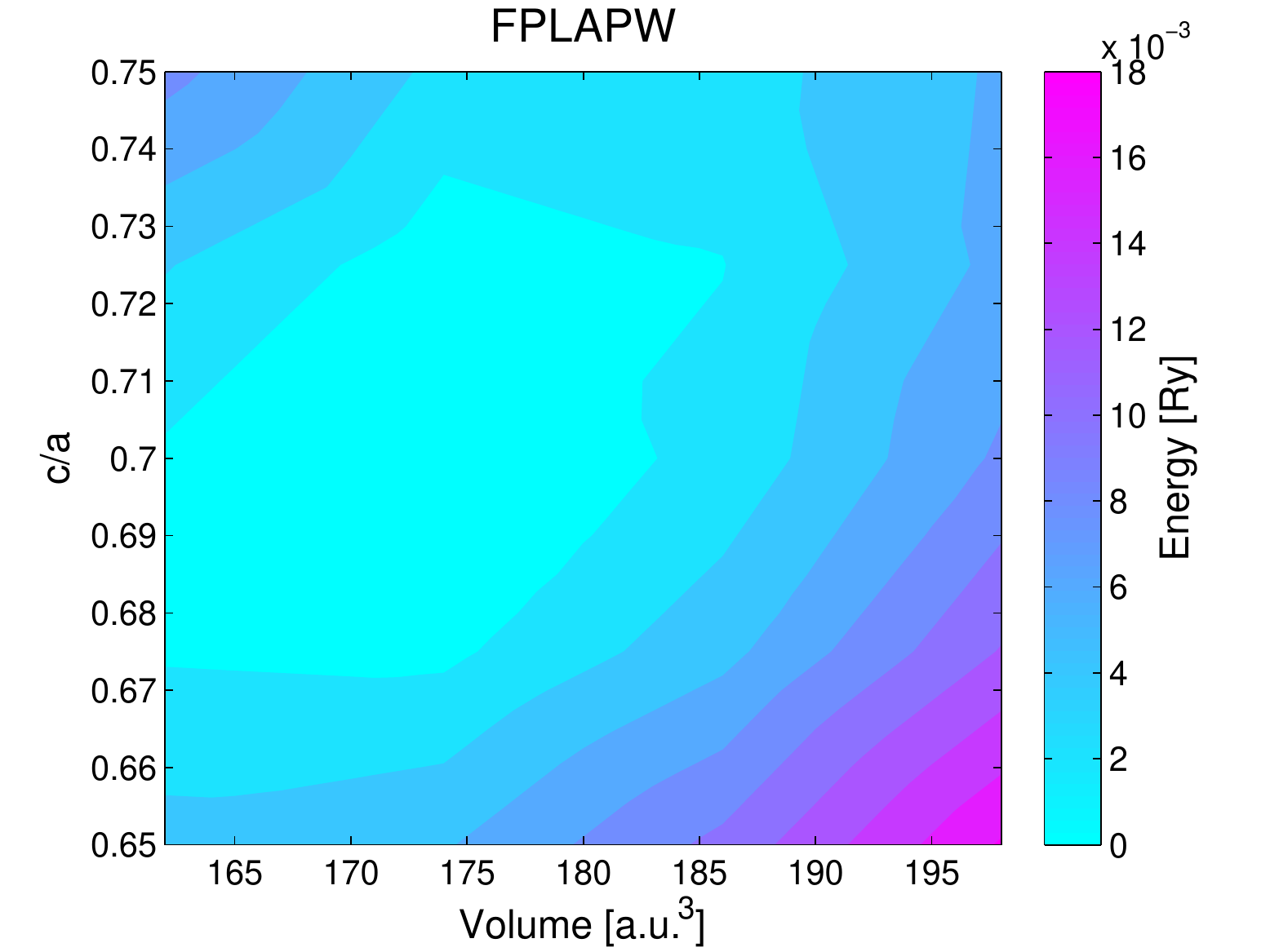}}
\caption{Left: Energy map as a function of volume and $c/a$-ratio for Li(2)$_{1.5}\square_{0.5}$Li(1)N, where the vacancy $\square$ is of type Li(2), calculated within EMTO-CPA. Right: Energy map as a function of volume and $c/a$-ratio for Li(2)$_2\square$N, where the vacancy is of type Li(1).}\label{fig_v5050p}
\end{figure}

In order to see how the lattice parameters change as the concentration
of vacancies increases, the structural disorder was modelled using the 
alloy analogy within the EMTO-CPA method. In this method the vacancy 
is treated as an `alloy component' represented by an empty atomic 
sphere. The system thus has the chemical formula Li(2)$_{2-x}\square_{x}$Li(1)N, 
where the vacancy is of Li(2) type. For $x=0.5$ one-fourth of the Li(2) atoms are
vacated. The results of the lattice parameter optimization performed within 
DFT calculations is presented in Fig.~\ref{fig_v5050p} (left). The equilibrium 
volume is $V_0 = 269.04$ a.u.$^3$ and the corresponding ratio $c/a = 1.065$. 
The EMTO results (Table~\ref{tab1}) for the clean Li$_3$N gives a lattice 
constant $a=3.538$ {\AA} while an in-plane Li(2)-vacated lattice Li(2)$_1\square_1$Li(1)N has 
$a=3.429$ \AA. From these two limits we propose the following linear 
concentration dependence of the lattice parameter: $a(x) = 3.429 + 0.109x$ {\AA}. 
For the concentration $x=0.5$ our total energy optimization gives for the 
lattice parameter $a$ the result: $a=3.483$ \AA. We note that this result agrees 
with the empirical Vegards-type law (linear increase of lattice constant as a 
function of concentration) to within $1\%$.

\subsection{Interplane vacancies at the Li(1)-site}
We next consider a full removal of the Li(1) atom, i.e. a complete removal
of the Li(1)-plane. For the practical calculation this corresponds to a crystal structure model in
which the Li(1)-plane consists of empty-spheres.
The energy landscape for the volume vs. $c/a$-ratio can be 
seen in Fig. \ref{fig_v5050p}. As the Li(1) atom is removed, the crystal 
becomes unstable and the $c/a$-ratio `collapses' to $\sim 0.7$. The decrease 
in the length of the $c$-axis is accompanied by a reduction in unit cell volume
to around $172.61$ a.u.$^3$, and consequently the spin-polarization vanishes. 
Due to this sizeable reduction of volume, $c/a$-ratio and magnetic moment, 
further computations with Li(1) vacancies was not carried on since it would lead to a 
significant departure from the structure of interest in experimental studies. Note
that a full removal of the Li(1)-plane was assumed in the calculations, and a single Li(1) vacancy will 
probably not result in such a large reduction of the $c/a$-ratio, and the magnetic moment 
might still remain in the case of a single vacancy.   

\subsection{Band structure}\label{res_bands}
\begin{figure}[h!]
\centering
\includegraphics[scale=0.5, clip=true]{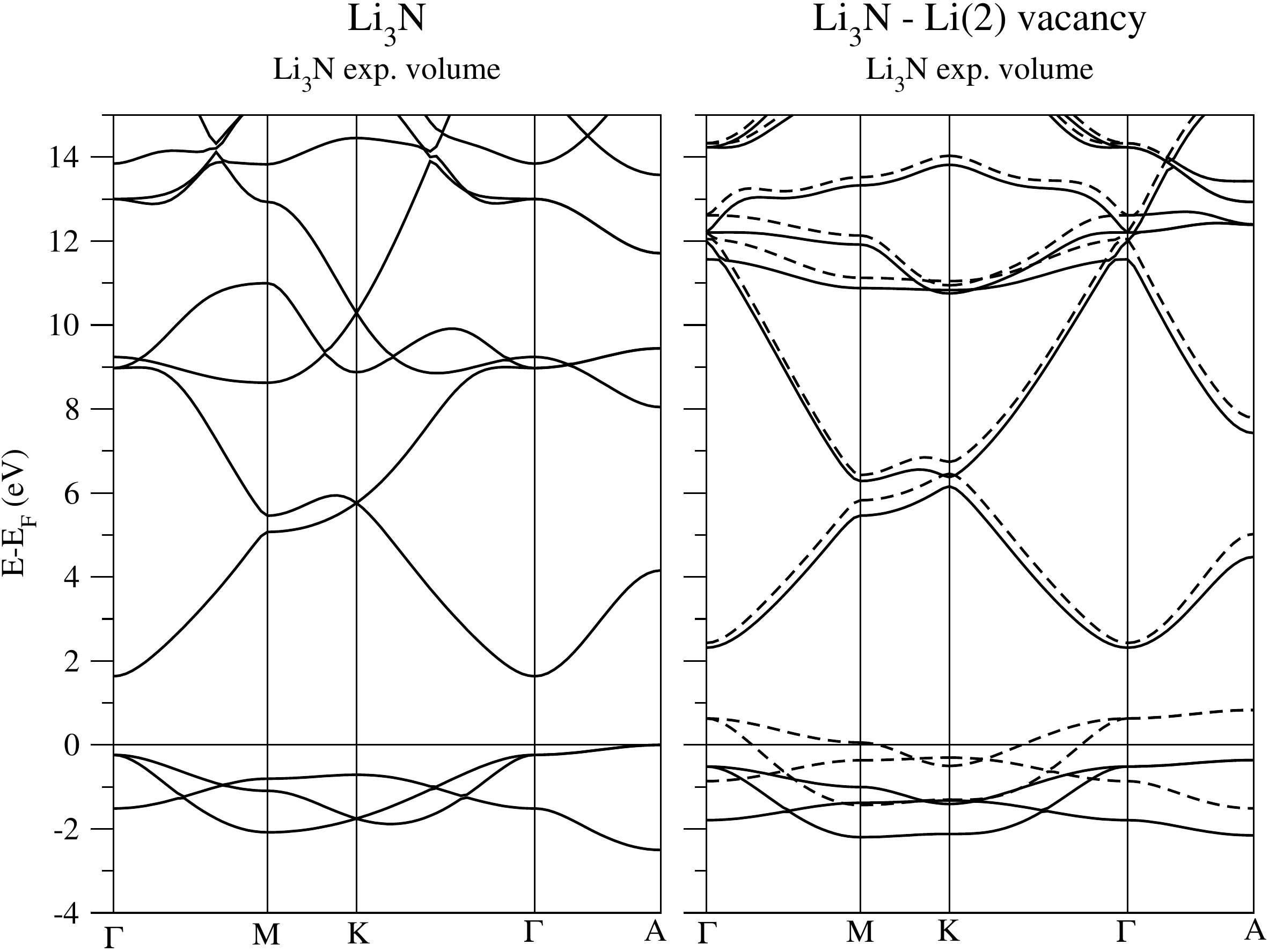}\\
\includegraphics[scale=0.5, clip=true]{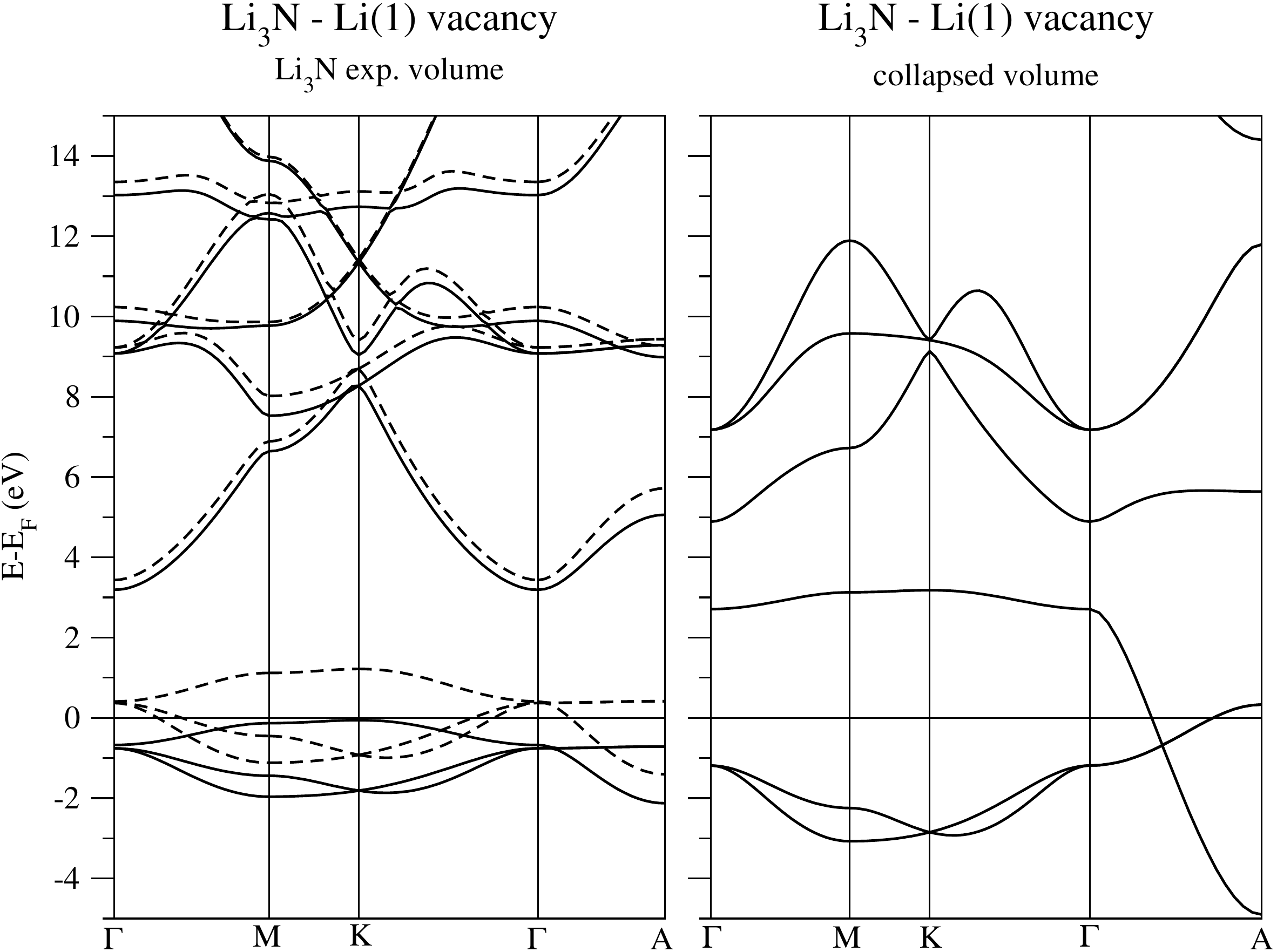}
\caption{Band structures. Top left: Perfect $\alpha-$Li$_3$N. Top right: Majority (full lines) and minority (dashed lines) bands for $\alpha-$Li$_3$N with one fully vacated Li(2) atom. Bottom left: Majority (full lines) and minority (dashed lines) bands for $\alpha-$Li$_3$N with one fully vacated Li(1) atom. Evaluated at the experimental volume of perfect Li$_3$N. Bottom right: Bands for the volume collapsed Li$_3$N with fully vacated Li(1) atom.}\label{bands}
\end{figure}

Results of the band structure calculations for the systems in discussion are shown
in Fig.~\ref{bands}. The band structure for perfect Li$_3$N, using the experimental
lattice parameters, is seen in 
the top left of Fig.~\ref{bands}, and agrees well with previous theoretical 
studies~\cite{kerk.81,la.yo.08}. The three valence bands have mainly N-$p$ orbital character, 
with the $p_x-$ and $p_y-$states being degenerate
at certain high symmetry points/directions within the Brillouin zone. The lowest
lying conduction band has mixed ($sp$) character, and its density is situated predominantly
in the region in between the Li$_2$N layers, as 
shown in Ref.~\cite{la.yo.08}. The gap ($\lesssim 1.5$ eV) underestimates the experimental
one ($\sim2.1$ eV~\cite{br.bl.79}), which is a common effect seen within DFT. Band
structure calculations within the PBE-GGA were also performed, and no major change in bands
and band gaps were found.

As a Li(2) atom is removed from the system, the N $p$-bands becomes partially occupied
and the chemical potential readjusts to account for the number of electrons (Fig.~\ref{bands}, top right).
The most drastic change is a spin-splitting of the bands: one spin channel being completely
occupied while the opposite one is almost rigidly shifted upwards in energy. 
Note that only the $p_x+p_y-$orbitals of the minority spin channel are shifted above
the Fermi level. In Fig.~\ref{bands} (bottom left) the results for the band structure 
for the perfect Li$_3$N cell with a completely removed Li(1) atom is seen. The result
is similar to the case of a Li(2) vacancy, namely: one spin channel is being shifted 
almost rigidly upwards in energy. In the case of Li(1) vacancy, we have identified the
$p_z-$orbital that is completely shifted above the Fermi level. In addition, the structural 
optimization showed that the system of widely separated Li$_2$N layers are
unstable, and that the structure collapses to a smaller volume and a smaller $c/a-$ratio. 
The band structure for the collapsed structure is shown in Fig.~\ref{bands} (bottom right). 
In this case, the remaining N-$p_z$ bands acquire a broad dispersion in the 
$\Gamma-A$-direction, suppressing the magnetism. We also performed calculations
for a pressurized system up to $20\%$ of the lattice constants 
(both isotropic and $c-$axial compression), and the system remained metallic.

\subsection{Magnetic ordering and vacancy concentration}\label{res_mag}

In this section we analyse the results of the electronic structure in 
view of the interplay between the magnetic moment formation and the vacancy 
concentration in Li$_3$N. In these calculations a constant volume equal to 
that of the Li(2)$_{1}\square_{1}$Li(1)N structure ($252.31$ a.u.$^3$) and the corresponding 
constant $c/a$-ratio equal to 1.07 were assumed.

\begin{figure}[h!]
\includegraphics[scale=0.35]{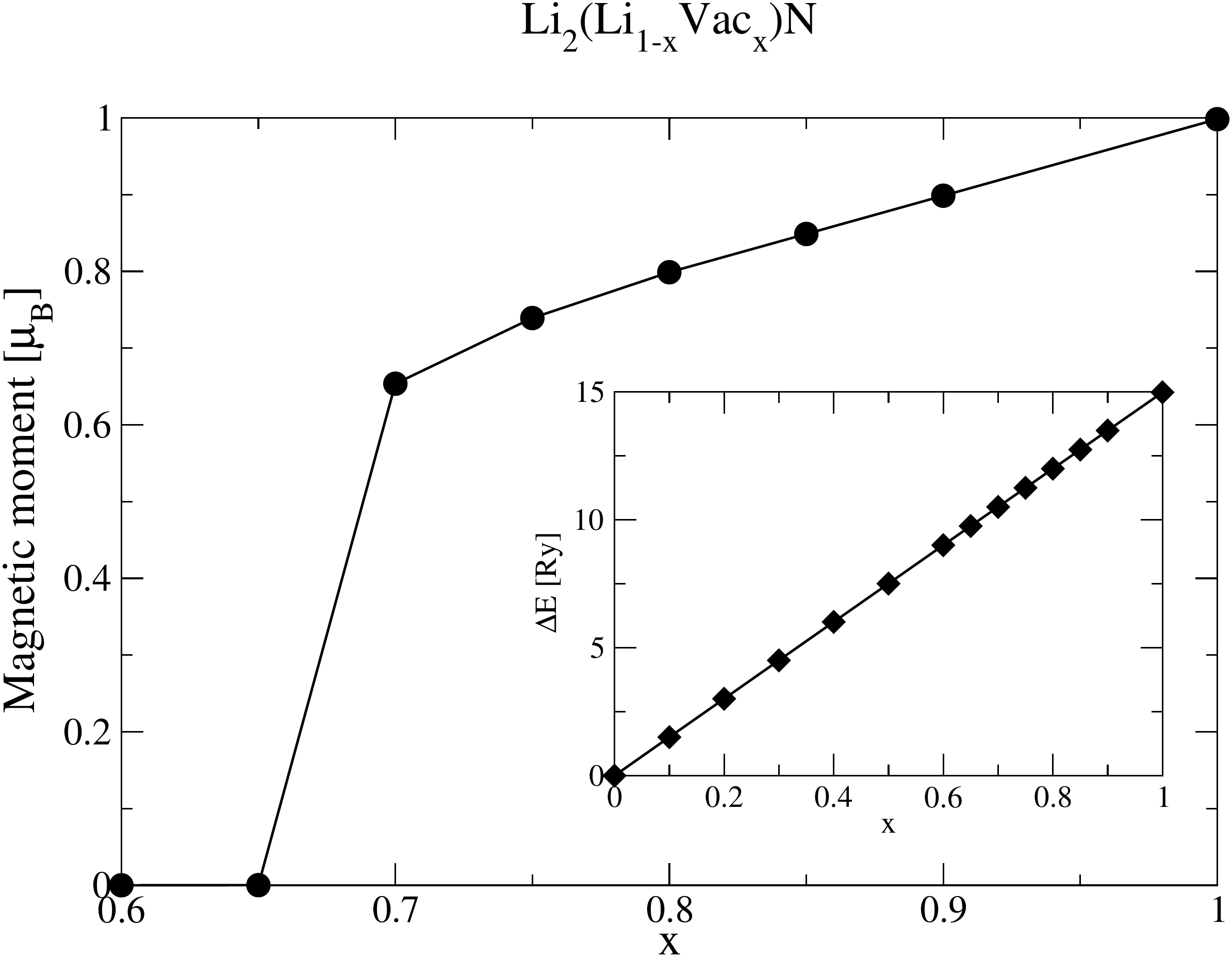}
\caption{Magnetic moment as function of concentration. Circles are calculated values, the solid lines are guides to the eye. Inset: Total energy difference $\Delta E$ between Li$_3$N and Li(2)$_{2-x}$Vac$_{x}$Li(1)N as a function of concentration $0 \leq x \leq 1$.}\label{magmom}
\end{figure}

For vacancy concentrations $x \leqslant 0.5$ no ferromagnetic order was formed.
For vacancy concentrations $x \geqslant 0.5$, the magnetic moment as a 
function of concentration can be seen in Fig. \ref{magmom}. 
Between $0.65 < x < 0.70$ a ferromagnetic onset occurs, with the magnetic 
moment being proportional to the vacancy concentration. At the complete vacancy 
substitution the moment is $\sim 1$ $\mu_B$.

We plot the difference between the pure Li$_3$N total energy and the total energy for the
Li(2)$_{2-x}\square_{x}$Li(1)N structure, in Fig.~\ref{magmom}. The energy difference
grows roughly linearly with increasing concentration, up to about $\sim 15$ Ry for the fully vacated
structure, which is expected for such a substantial structural change (one fully removed Li(2) atom). 
For an estimate of the formation energy a supercell approach with relaxation
of the atomic positions should be performed~\cite{wu.do.09,de.de.09,li.zh.14,li.zh.15}, which is outside the scope of the present study.

\begin{figure}[h!]
\centering
\includegraphics[scale=0.4]{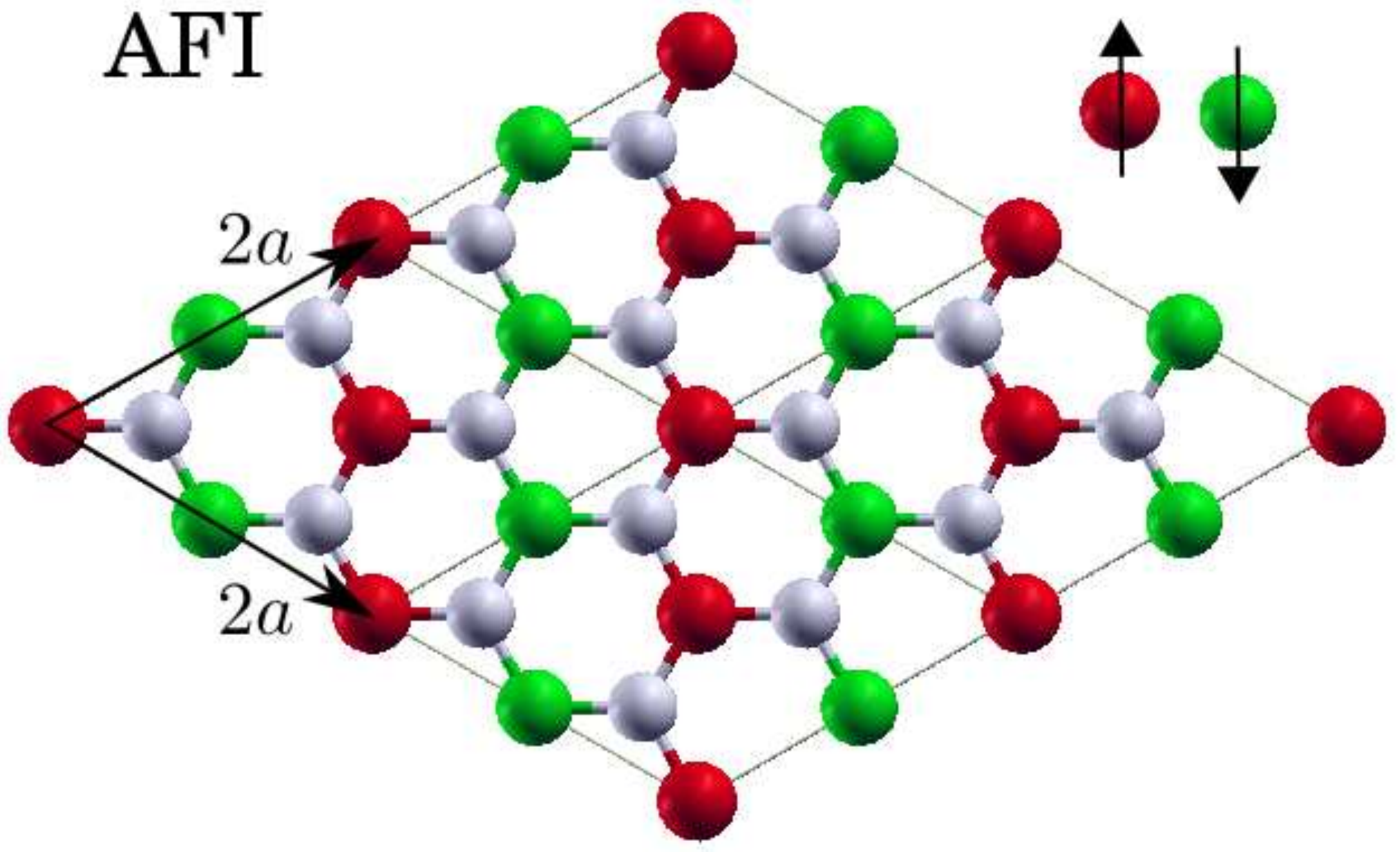}\includegraphics[scale=0.3]{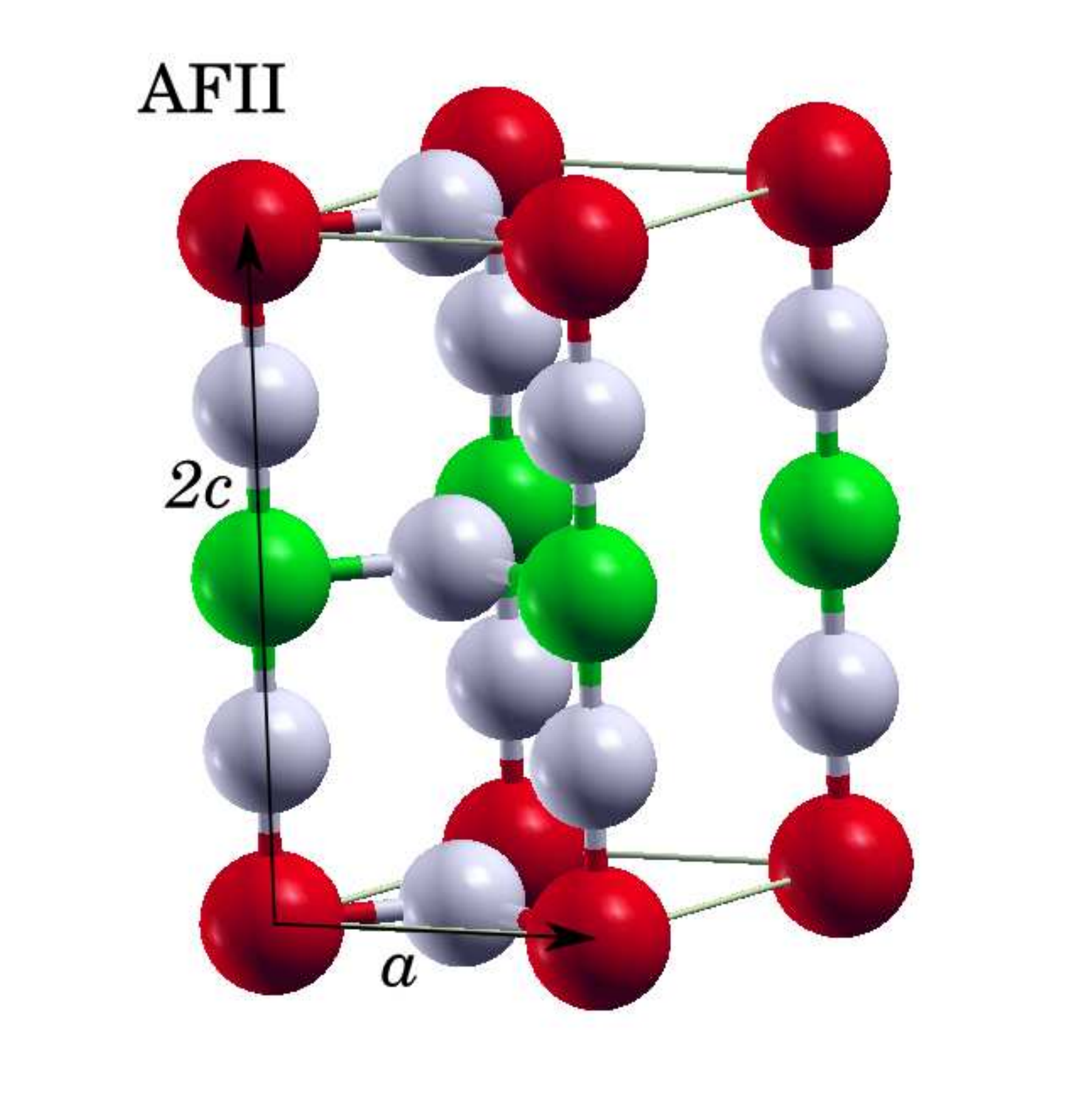}
\caption{AFI/II-type collinear magnetic state (left/right). The magnetic moment points along
the $c-$axis; red (up); green (down). Grey spheres correspond to Li.}\label{antiferro}
\end{figure}

In order to determine the ground state magnetic order present in the system, several calculations were performed
assuming different antiferromagnetic (AFM) structures, as presented in Fig. \ref{antiferro}.
The antiferromagnetic supercells were calculated with the LMTO-ASA method~\cite{ande.75}.
Two configurations were investigated, see Fig.~\ref{antiferro}. One configuration with antiferromagnetic ordering
\emph{within} the basal $ab-$plane (AFI, Fig.~\ref{antiferro} (left)), and one where the moments \emph{between}
the basal planes couple antiferromagnetically (AFII, Fig.~\ref{antiferro} (right)).

For all studied AF structures, the total energies were higher than that of the 
FM solution (magnetic ground state configuration), but lower than the non-magnetic (NM) 
solution, $E_{FM}<E_{AFM}<E_{NM}$. 
For the AFM ordering with the moments alternating within the $ab$-plane 
(AFI, Fig. \ref{antiferro}, left), the energy differences with respect to the
ferromagnetic ground state is estimated to be $|\Delta E_{FM-AFM}|=2.4$ mRy/atom ($\approx 370$~K).  
The non-magnetic state is situated at $|\Delta E_{FM-NM}|=2.6$ mRy/atom ($\approx 400$~K)
above the ferromagnetic groundstate. 
For the AFM ordering with the moments alternating between consecutive layers (AFII, Fig. \ref{antiferro}, right) 
the energy difference was $|\Delta E_{FM-AFM}|=0.24$ mRy ($\approx 40$ K), an order of magnitude smaller than the AFI configuration. Hence,
the magnetic coupling between neighbouring Li$_2$N layers can be considered negligible compared to the intralayer coupling.

\section{Discussion}\label{sec_discus}
Our results based on the DFT calculations show that the complete removal 
of one Li atom (in plane or between planes) triggers the magnetic 
moment formation, which is a consequence of the partial occupation 
of the N $p-$orbital. In the case of a full removal of the (out-of-plane)
Li(1) atom, the crystal structure is unstable and would undergo a large reduction 
in volume. If one of the geometrically equivalent Li(2) atoms is removed, 
the volume reduction is less drastic. 
To our best knowledge no experimental studies exist in which the full Li(2) 
vacancy has been achieved.
On the other hand, in the LiNiN structure~\cite{ba.bl.99,st.go.04} one of the Li(2) 
positions has been vacated and at the same time
the Li(1) atom has been fully substituted by Ni. 
In this case, as expected the Ni $d-$bands are 
situated at the Fermi level, and thus the problem gains complexity,
since correlation physics has to be included. Such calculations
are outside the scope of the present study but shall be considered in the future.

In our Li(2) vacated structure, the N binds 3 lithium ions within the basal plane. 
A similar local environment around the nitrogen is realized in the basal plane of
$\beta-$Li$_3$N, which can be reached from $\alpha-$Li$_3$N by applying
modest pressure ($\sim 0.5$ GPa)~\cite{la.ma.05}. At this values of pressure,
three of the six bonds of $\alpha-$Li$_3$N are broken leading to 3 Li nearest neighbours 
around the N atom.

From our CPA calculations, the vacancy concentration leading to moment formation can be estimated. 
Early studies showed that the natural vacancy concentration of Li$_3$N
is about $1-2\%$ at the Li(2) positions, and that this does not change significantly
with increasing temperature~\cite{sc.th.79}. On the other hand, vacancies can also be created in a 
non-equilibrium situation, similarly to the one existing in battery cells where a voltage has been applied.
The ionic conductivity will then remove Li from the system, and a larger concentration
of vacancies may be achieved. Therefore, further experimental studies of magnetic 
properties as a function of vacancy concentration are desired to verify
our theoretical estimations presented in this paper.

\section{Conclusions}\label{sec_conc}
In this work we have performed density functional theory calculations 
for Li vacancies in $\alpha-$Li$_3$N. Assuming unchanged hexagonal crystal 
symmetry, and a complete removal of a Li(1) atom 
we showed by lattice relaxation that the unit cell volume and 
$c/a-$ratio will be drastically reduced and no magnetic moment 
is formed. By vacating the in-plane Li(2) atom for amounts larger than $>65\%$ 
we predict that $\alpha-$Li$_3$N can be turned metallic and ferromagnetic. 
The removal of 
a Li(2) atom will cause a less drastic structural reduction, and a magnetic 
ground state sets in with an ordered ferromagnetic moment of 1 $\mu_B$.

\section*{Acknowledgements}
The authors would like to acknowledge Anton Jesche for helpful discussions.
The Deutsche Forschungsgemeinschaft through TRR80 is gratefully acknowledged for financial support.
We acknowledge computational resources provided by the Swedish National Infrastructure for Computing (SNIC) at the National Supercomputer Centre (NSC) in Link\"{o}ping.


\end{document}